\documentclass[utf8]{FrontiersinHarvard}

\usepackage{url,hyperref,lineno,microtype}
\usepackage[onehalfspacing]{setspace}
\usepackage{booktabs}

\def\keyFont{\fontsize{8}{11}\helveticabold }
\def\firstAuthorLast{Ivanov {et~al.}}
\def\Authors{Vasily Ivanov\,$^{1,*}$, Polina Verezemskaya\,$^{2}$, Alexander Gavrikov\,$^{2}$, Vitaliy Sharmar\,$^{2}$, Mikhail Krinitskiy\,$^{3,2}$, Timofey Grigoryev$^{1}$, Vladimir Vanovskiy\,$^{1}$ and Evgeny Burnaev\,$^{1,4}$}
\def\Address{$^{1}$Artificial Intelligence Center, Skolkovo Institute of Science and Technology, Moscow 121205, Russia \\
$^{2}$Shirshov Institute of Oceanology, Russian Academy of Sciences, Moscow, Russia \\
$^{3}$Moscow Institute of Physics and Technology, Moscow, Russia \\
$^{4}$Autonomous Non-Profit Organization Artificial Intelligence Research Institute (AIRI), Moscow 105064, Russia}
\def\corrAuthor{Vasily Ivanov}
\def\corrEmail{\detokenize{va.ivanov@skoltech.ru}}

\begin{document}
\onecolumn
\firstpage{1}

\title[Borey forecasting system and dataset]{Borey: A High-Resolution Regional Atmosphere--Ocean--Sea Ice--Wave Forecasting System and Hindcast Dataset for the Barents and Kara Seas}

\author[\firstAuthorLast ]{\Authors}
\address{}
\correspondance{}
\extraAuth{}
\maketitle

\begin{abstract}
\section{}
Borey is a high-resolution regional modeling and operational forecasting system for the Barents and Kara Seas. It combines WRF for the atmosphere, NEMO-SI3 for the ocean and sea ice, and WW3 for waves on approximately 3--6\,km grids, and generates daily forecasts to 72 hours. We describe the model chain and production workflow and present an accompanying hourly hindcast of surface conditions from August 2015 to August 2023. The archive provides aligned atmosphere, ocean, sea ice, and wave fields for regional marine studies and a baseline for evaluating the operational system. Comparisons with observations and observation-based products show that Borey captures much of the variability in near-surface atmospheric conditions and ocean temperature. Skill in the evaluated WRF, NEMO, and SI3 forecasts changes only modestly across the three-day window. The main limitations are persistent rather than rapidly growing errors: sea surface temperature is generally too cold, sea ice concentration and occurrence are overestimated during seasonal retreat, and significant wave height is underestimated. Borey should therefore complement observation-constrained products. The planned public release will provide the complete hourly surface dataset on its native grids, together with configuration and provenance documentation, for regional analysis, model development, and carefully evaluated data-driven forecasting and data-assimilation research.

\tiny
\keyFont{ \section{Keywords:} Arctic hindcast, Barents Sea, Kara Sea, sea ice, regional ocean modeling, wave modeling, marine dataset}
\end{abstract}

\section{Introduction}

The Barents and Kara Seas are rapidly changing Arctic shelf seas with pronounced spatial and seasonal variability \citep{arctic_change_review,barents_kara_context}. Marine operations, environmental assessment, and process studies require consistent information on weather, ocean state, sea ice, and waves, particularly near coasts and the marginal ice zone. Global analyses and forecasts provide essential large-scale context, but regional applications also require a system that resolves shelf geometry and maintains consistent fields across components.

Relative to commonly used global atmospheric and wave products, Borey's 3--6\,km component grids provide finer regional sampling of shelf and coastal variability \citep{era5_dataset,gfs_dataset,waverys_dataset}. Its ocean resolution is comparable to high-resolution products such as GLORYS12 and to regional Arctic systems such as TOPAZ5, rather than uniformly finer than every available product \citep{glorys_dataset,topaz_dataset}. Borey's contribution is the combination of an aligned hourly multicomponent archive with the documented operational chain that produces daily short-range forecasts.

Regional hindcasts can complement assimilative products by providing internally consistent multivariable records without analysis increments. Their utility depends on more than nominal resolution: forcing, open boundaries, restart sequencing, archive structure, and validation must also be documented. A free-running trajectory is valuable for model evaluation and downstream experiments, but it can retain or accumulate systematic error because the interior state is not corrected by observations.

This paper presents Borey as both an operational atmosphere--ocean--sea ice--wave system and the source of an hourly regional hindcast. Borey combines WRF, NEMO-SI3 in the BARKA12 configuration, and WW3 \citep{wrf_model,nemo_model,si3_model,ww3_model}. We document the model chain, production workflow, and surface dataset; evaluate the hindcast against observations and observation-based products; and retain model-based comparisons as contextual benchmarks. We also assess lead-time skill for the WRF, NEMO, and SI3 operational forecasts. Wave and surface-current forecast validation remains in progress.

\section{Materials and methods}

\subsection{Domain and model chain}

\begin{figure}[t]
  \centering
  \includegraphics[width=0.88\linewidth]{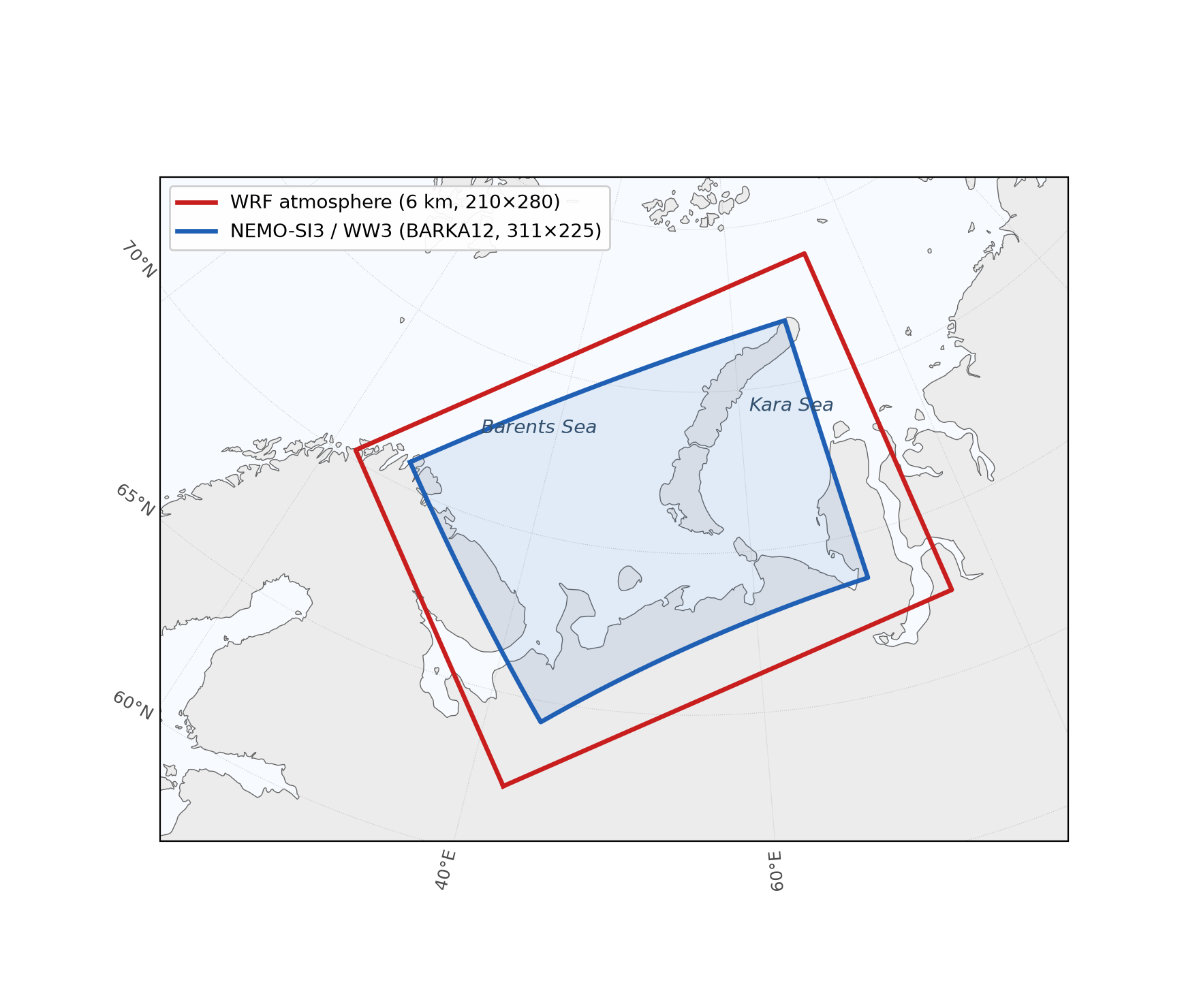}
  \caption{Borey model domains. The blue outline shows the NEMO-SI3/BARKA12 ocean--sea ice and WW3 grid; the red outline shows the larger WRF atmospheric domain.}
  \label{fig:domain}
\end{figure}

The Borey ocean--sea ice domain covers the southern Barents Sea, the Kara Sea, and adjacent shelf waters (Figure~\ref{fig:domain}). BARKA12 uses a $311\times225$ curvilinear grid at nominal $1/12^{\circ}$ resolution, corresponding to approximately 3--5\,km across the domain, with 40 vertical levels. Its open boundaries connect the regional model to the large-scale fields used for initialization and boundary forcing.

The production chain is sequential and one-way. WRF supplies atmospheric fields to NEMO-SI3 and WW3, while NEMO-SI3 supplies the ocean and sea ice fields required by the wave configuration. Components exchange prepared files rather than fluxes through an online coupler. This design supports reproducible restarts and independent component reruns but does not represent two-way atmosphere--ocean--wave--ice feedbacks.

\subsection{Component configurations}

The atmospheric component is based on the non-hydrostatic Weather Research and Forecasting (WRF) Model, version 4.4.2 \citep{wrf_model,powers2017}. The computational domain (Figure~\ref{fig:domain}, red box) has a horizontal grid spacing of 6\,km, 50 vertical levels (approximately 20 below 850\,hPa), and hourly output. The model is initialized from the National Centers for Environmental Prediction (NCEP) Global Forecast System (GFS) operational analyses at $0.25^{\circ}$ spatial and 6-hourly temporal resolution, which also provide the lateral boundary conditions \citep{gfs_dataset}.

Turbulent exchange within the atmospheric boundary layer is represented by the Mellor--Yamada--Nakanishi--Niino (MYNN) planetary boundary layer scheme \citep{nakanishi2009}, coupled with the corresponding MYNN surface-layer parameterization for a physically consistent treatment of surface fluxes. Land-surface processes use the Noah Land Surface Model \citep{chen2001}, and cloud microphysics use the WRF Single-Moment 6-class (WSM6) scheme \citep{hong2006}. Deep convection is explicitly resolved, so no cumulus parameterization is employed. Longwave and shortwave radiative transfer is computed with the Rapid Radiative Transfer Model for General Circulation Models (RRTMG) \citep{iacono2008}, using the monthly climatological ozone and CAM aerosol options.

To improve the representation of Arctic surface conditions, fractional sea ice coverage is enabled so that sub-grid sea ice concentration contributes to the surface turbulent and radiative fluxes, which are computed interactively by the MYNN surface-layer scheme. Maximum snow albedo is prescribed from the static geographical dataset, deep soil temperature is updated throughout the simulations to account for seasonal subsurface variations, and topographic wind correction is enabled over complex terrain. This physics configuration is designed to represent stable Arctic boundary layers, air--surface interaction over snow- and ice-covered surfaces, and mesoscale atmospheric processes; comparable WRF-based configurations have been used in recent Arctic modeling studies \citep{bertossa2026,zhang2026}.
Native grid-relative winds are rotated to geographic eastward and northward components before validation or downstream use.

The ocean--sea ice component is based on NEMO 4.0.7 and SI3 in the regional BARKA12 configuration \citep{nemo_model,si3_model}. BARKA12 has a $311\times225$ curvilinear grid and 40 fixed $z$ levels with partial bottom cells. The model uses a 300\,s baroclinic time step, a linear free surface, and a split-explicit barotropic solver. Seawater properties follow TEOS-10. Tracer transport employs fourth-order flux-corrected advection and grid-scaled isoneutral Laplacian diffusion. Momentum is represented using third-order upstream-biased advection, an energy- and enstrophy-conserving vorticity scheme, and grid-scaled horizontal bilaplacian viscosity. Vertical mixing is parameterized with a turbulent kinetic energy closure that includes Langmuir-cell and enhanced-convection terms; a diffusive--advective bottom boundary layer and nonlinear implicit bottom drag are also applied.

Surface fluxes are calculated at every ocean time step with the NCAR bulk formulation using WRF fields interpolated to the BARKA12 grid. The forcing comprises 10\,m winds, 2\,m air temperature and humidity, precipitation, snowfall, mean sea-level pressure, and downwelling shortwave and longwave radiation. Surface stress includes a current-feedback correction and a concentration-dependent air--ice drag coefficient. Penetrative shortwave radiation is represented with the two-band scheme, with a diurnal cycle reconstructed from the daily shortwave forcing. Monthly climatological river discharge is prescribed, together with enhanced mixing over the upper 10\,m at river mouths. Tidal forcing, the atmospheric-pressure contribution to ocean dynamics, ice-shelf processes, and wave-induced ocean forcing are not included.

For a cold start, ocean temperature and salinity are initialized from daily GLORYS12 fields. Daily GLORYS12 sea surface height, velocity, temperature, salinity, sea ice concentration, and sea ice thickness are also imposed at the southern, western, and northern open boundaries \citep{glorys_dataset}. Flather conditions are used for the barotropic flow, while three-dimensional velocity, tracers, and sea ice use flow-relaxation conditions. Temperature and salinity are not restored within the model interior. Surface salinity is restored toward daily GLORYS12 values after spatial filtering, with the restoring flux scaled by the open-water fraction. No data assimilation is used. After initialization, consecutive production cycles are continued from ocean and sea ice restart files.

SI3 dynamics and thermodynamics are advanced every 300\,s. BARKA12 uses a single ice-thickness category with the virtual ice-thickness-distribution parameterization, rather than the five explicit categories in the SI3 reference configuration, together with two ice layers and one snow layer. Ice dynamics use non-adaptive elastic--viscous--plastic rheology with 120 subcycles, Prather advection, Hibler-type ice strength, and explicit ridging and rafting. The ice--ocean drag coefficient is 0.01, and the optional landfast-ice parameterization is disabled. Thermodynamics include Bitz--Lipscomb heat diffusion, Pringle thermal conductivity, variable ice salinity, open-water ice formation, lateral melting, and level-ice melt ponds with frozen lids and albedo feedback; heat supplied to leads is used to melt ice before warming the ocean. The Lebrun shortwave-transmission scheme is used, with a minimum floe diameter of 7\,m for lateral melting. At a cold start, sea ice is diagnosed from sea surface temperature using nominal Northern Hemisphere values of 1\,m for thickness and 0.7 for concentration; subsequent cycles use restart fields, while GLORYS12 sea ice concentration and thickness continue to constrain the open boundaries.

The wave component uses WAVEWATCH III (hereafter WW3), forced by hourly WRF 10\,m winds and the NEMO-SI3 sea ice concentration \citep{ww3_model}. A curvilinear grid sharing the BARKA12 domain configuration ($311\times225$) is used to simulate wave propagation from the Barents Sea through the Kara Gate to coastal waters; reusing the common ocean--sea ice grid minimizes interpolation error in a region of rapidly varying ice conditions. The spectrum has 36 directions and 30 frequencies spanning approximately 0.0418 to 0.7293\,Hz. The four model time steps are $\Delta t_{g}=210$\,s (global), $\Delta t_{xy}=55$\,s (spatial advection), $\Delta t_{k}=105$\,s (intra-spectral), and $\Delta t_{s}=15$\,s (source terms). Wind input and dissipation use the ST4 source terms \citep{ardhuin2010semiempirical}, nonlinear interactions use the Discrete Interaction Approximation (DIA), and wave--ice interaction uses the IC0 scheme.
The published wave variables include significant wave height and peak frequency; wave period is derived as the inverse of peak frequency where used.

\begin{table}[t]
\centering
\caption{Borey components and archived paper coverage used in this study. August 2015 is the common dataset start; reference-specific validation overlaps may begin later.}
\label{tab:model_config}
\resizebox{\textwidth}{!}{%
\begin{tabular}{llllll}
\toprule
Component & Model & Grid & Evaluated period & Cadence & Principal variables \\
\midrule
Atmosphere & WRF 4.4.2 & $\sim$6\,km, $210\times280$ & 2015-08 to 2023-08 & Hourly & 10\,m wind, 2\,m temperature, pressure, humidity, fluxes \\
Ocean & NEMO 4.0.7 & BARKA12, $311\times225$, 40 levels & 2015-08 to 2023-08 & Hourly/daily & Temperature, salinity, sea level, currents, mixed-layer depth \\
Sea ice & SI3 & BARKA12 & 2015-08 to 2023-08 & Hourly & Concentration, thickness, drift, snow and stress diagnostics \\
Waves & WW3 & BARKA12-domain wave grid & 2015-08 to 2023-08 & Hourly & Significant wave height, peak frequency, stresses \\
\bottomrule
\end{tabular}%
}
\end{table}

\subsection{Production pipeline and computational environment}

\begin{figure}[t]
  \centering
  \includegraphics[width=0.95\linewidth]{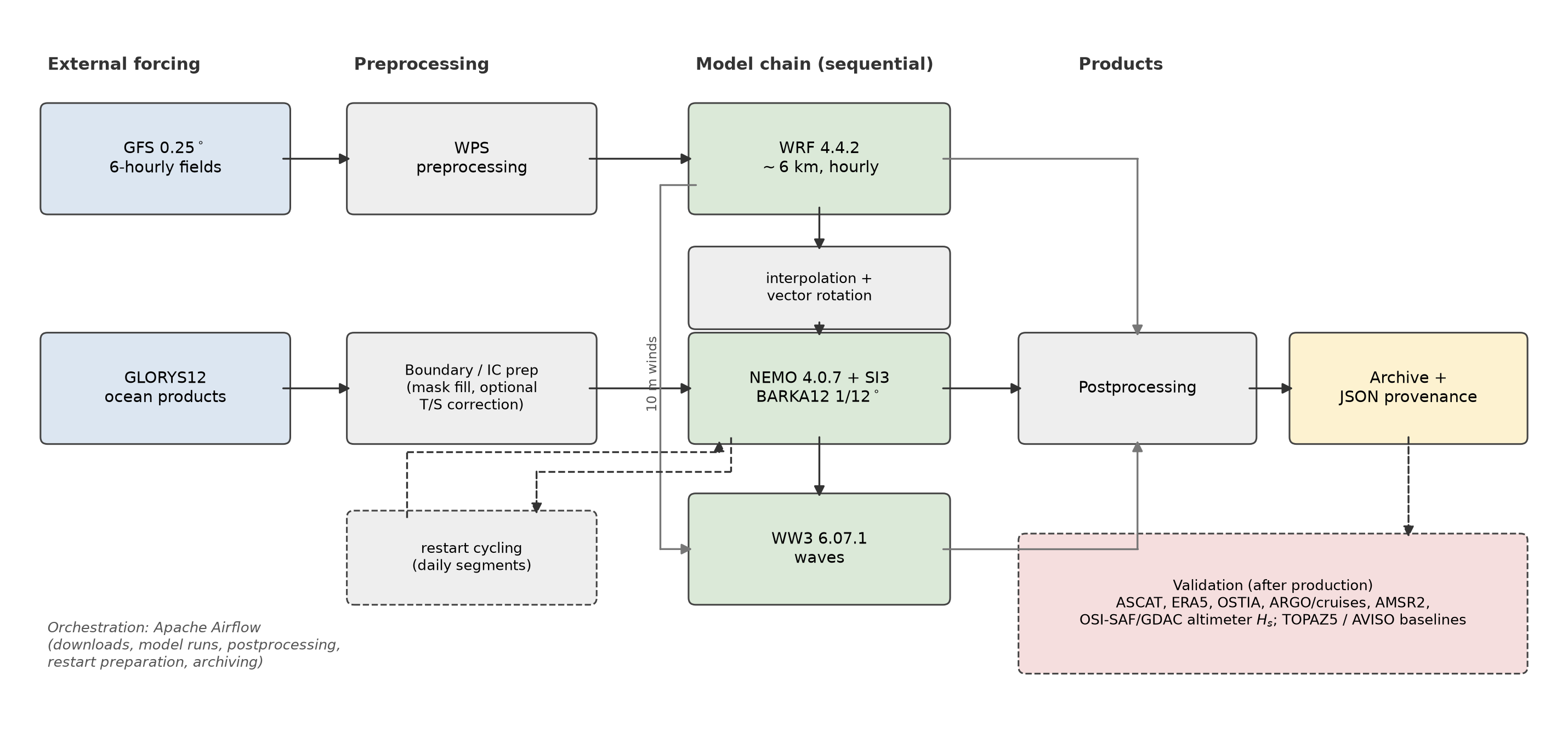}
  \caption{Borey production workflow. External forcing is prepared for the sequential WRF, NEMO-SI3, and WW3 runs; outputs, restarts, logs, and provenance records are then archived. Validation is performed after production.}
  \label{fig:workflow}
\end{figure}

Apache Airflow orchestrates forcing acquisition, preprocessing, model execution, postprocessing, restart preparation, and archiving (Figure~\ref{fig:workflow}). The hindcast is stored internally as \texttt{reanalysis\_no\_ass}; despite this directory name, it is free-running and assimilates no ocean or sea ice observations. The paper record extends from August 2015 through August 2023.

Component configuration files, forcing manifests, restart inputs, and time-stamped workflow logs define each production cycle. Airflow checks upstream completion before launching the next component and stages restart and output files separately, allowing failed tasks to be rerun without silently replacing a completed cycle.
The internal production record associates each cycle with versioned model configurations, source-code revisions, and forcing and restart manifests.

All production components were executed sequentially on a single x86\_64 compute node built on AMD EPYC 7763 64-Core Processor hardware, exposing 128 logical CPUs and equipped with 512\,GB RAM. The models were built with the 2021 release of the Intel compiler suite. For a 72-hour forecast cycle, the mean wall-clock runtimes were 1\,h\,28\,min for WRF, 8\,min for NEMO-SI3, and 50\,min for WW3.

NEMO-SI3 output is stored in daily cycles containing an initial 24-hour segment followed by a 72-hour forecast. The continuous hindcast used for validation is assembled from the first 24 valid hours of successive 00 UTC cycles after decoding each file's time coordinate. Overlapping later hours are excluded from the hindcast and retained only for lead-time diagnostics. WRF and WW3 use their native hourly coordinates and component-specific coverage.

The public release will contain the complete hourly surface dataset, its native grids and masks, and documentation of the model configurations and production provenance. Three-dimensional NEMO fields are used only to assess subsurface temperature and salinity in this study and will not be distributed. Release documentation will specify the data format, metadata conventions, repository, persistent identifier, license, and inventory of any missing periods.

\subsection{Validation design}
\subsubsection{Hindcast validation}

Validation uses observations and observation-based products where available and distinguishes them from forcing products and contextual model comparisons (Table~\ref{tab:validation_design}). The principal references are ASCAT, OSTIA, Argo and local cruises, AMSR2 JAXA, and ESA CCI Sea State \citep{ascat_dataset,ostia_dataset,argo_dataset,jaxa_amsr2_dataset,cci_sea_state_dataset}. ERA5 provides a spatially complete atmospheric reanalysis benchmark \citep{era5_dataset}. GFS and GLORYS are not used as independent validation because they drive or constrain Borey. TOPAZ5 is retained as an assimilative model benchmark \citep{topaz_dataset}, and AVISO/DUACS represents geostrophic rather than total surface flow.

\begin{table}[t]
\centering
\caption{Primary Borey validation comparisons. Exact product versions, quality-control flags, masks, and collocation tolerances are recorded in the corresponding metric specifications.}
\label{tab:validation_design}
\resizebox{\textwidth}{!}{%
\begin{tabular}{lllll}
\toprule
Component & Variable & Reference & Reference role & Evaluated period \\
\midrule
WRF & 10\,m wind speed & ASCAT & Satellite observation & 2019-01 to 2023-08 \\
WRF & 10\,m wind; 2\,m air temperature & ERA5 & Reanalysis benchmark & 2015-08 to 2022-10 \\
NEMO & Sea surface temperature & OSTIA & Observation-based analysis & 2015-08 to 2023-08 \\
NEMO & Conservative Temperature and Absolute Salinity profiles & Argo and local cruises & In situ observations & 2020-03 to 2023-07 \\
SI3 & Sea ice concentration and ice edge & AMSR2 JAXA & Satellite observation & 2015-10 to 2023-08 \\
WW3 & Significant wave height & ESA CCI Sea State Level-3 & Satellite observation & 2015-08 to 2023-08 \\
WW3 & Significant wave height & CMEMS ALLSAT Level-4 & Satellite observation product & 2020-01 to 2023-08 \\
\bottomrule
\end{tabular}%
}
\end{table}

The primary independent wave reference is ESA CCI Sea State version 4 Level-3 significant wave height \citep{cci_sea_state_dataset}. The source product combines calibrated along-track radar-altimeter observations from multiple missions and provides adjusted and empirically denoised estimates \citep{huang1998empirical,quilfen2019denoising}. The selected record is identified by CEDA UUID \texttt{82273fe59b784e7c914e1bd3049e3483}. For the present workflow, CCI is available locally as complete monthly hourly files on a 0.25$^{\circ}$ grid, and the comparison uses \texttt{swh\_denoised}. The configured period is August 2015 through August 2023. Quality control requires finite, unmasked \texttt{swh\_denoised}, \texttt{count} $>$ 0, and any declared valid range; matchup metrics are not weighted by \texttt{count}.

A complementary comparison uses the CMEMS ALLSAT multi-mission Level-4 significant-wave-height product, represented locally as nominally hourly monthly files on a 0.5$^{\circ}$ grid. Missing reference timestamps are omitted rather than interpolated; the workflow permits at most 24 missing hours in any month and records each omission. Product identifiers and processing provenance for this local hourly representation will accompany the validation release. The comparison covers January 2020 through August 2023. For every finite, non-negative ALLSAT value, WW3 \texttt{hs} is sampled at the nearest model-grid center within 15\,km and 30\,min. Matchups with SI3 sea ice concentration of at least 0.15 are excluded. Because the CCI and CMEMS products draw on overlapping multi-mission altimeter records, they are treated as complementary data representations rather than statistically independent observing systems.

Gridded surface products are compared after common valid-data and domain masks are applied. Sampling or aggregation to the comparison grid follows each metric specification. For the local CCI derivative, WW3 \texttt{hs} is collocated only with occupied CCI cells, and the SI3 ice mask is applied before matchup arrays are written. Argo and cruise observations are matched to the nearest jointly wet model column within 15\,km; model Conservative Temperature and Absolute Salinity are then interpolated to observed depths without vertical extrapolation. Product-specific tolerances are used for native along-track observations. Vector reference frames are harmonized before scoring.

Scalar comparisons report sample count, mean model-minus-reference bias, mean absolute error (MAE), root-mean-square error (RMSE), and Pearson correlation. Sea ice concentration validation also counts grid-cell ice-state disagreements at the 0.15 concentration threshold and separates model-only ice from reference-only ice. Vector comparisons report speed statistics, root-mean-square vector error, complex vector correlation \citep{crosby1993}, and absolute and signed angle errors. The accompanying validation protocol records formulas, masks, collocation tolerances, reference-frame rotations, and metadata checks.

\subsubsection{Forecast validation}
\label{sec:forecast_validation}

As a diagnostic of the operational chain, we evaluate complete 00 UTC forecast cycles at lead times of 24, 48, and 72 hours. Two-metre air temperature and 10\,m wind speed are compared with ERA5, sea surface temperature with OSTIA, and sea ice concentration and ice edge with AMSR2 JAXA. Each field is matched at its valid time using the masks, quality control, units, and spatial rules of the corresponding hindcast metric. Within each comparison, only cycles containing all three leads enter the common-cycle cohort. The evaluated cycles come from operational production between July 2025 and February 2026; cohort sizes differ among components because their forecast availability differs.

The forecast diagnostic reports the same scalar scores as the hindcast evaluation. It is intended to show change with lead within a common cohort, not to compare absolute forecast and hindcast skill: the recent operational cycles cover a different period and season from the multi-year hindcast. Significant-wave-height and surface-current forecast validation is not yet complete and is therefore omitted.

\section{Results}

\subsection{Dataset coverage}
The evaluated Borey record comprises four component archives from August 2015 through August 2023. The planned dataset contains hourly surface variables on their native grids. Daily three-dimensional NEMO fields remain in the internal production archive; they support the profile validation reported here but are not part of the release. Table~\ref{tab:archive_shapes} summarizes representative internal dimensions and variables. The release inventory will give exact first and last valid times, gaps, and the final surface-variable list.

\begin{table}[t]
\centering
\caption{Representative dimensions and variables from the internal Borey archive used in this study. The public release inventory will provide the exhaustive variable list and metadata for the hourly surface dataset. NEMO-SI3 uses $(x,y)=(311,225)$, whereas the WW3 files store the horizontal axes as $(\mathrm{latitude},\mathrm{longitude})=(225,311)$.}
\label{tab:archive_shapes}
\resizebox{\textwidth}{!}{%
\begin{tabular}{llll}
\toprule
Archive & Representative shape & Cadence/file organization & Variables used \\
\midrule
WRF native fields & $24\times210\times280$ & Hourly; daily files & \texttt{U10}, \texttt{V10}, \texttt{T2}, \texttt{Q2} \\
NEMO surface & $24/72\times311\times225$ & Hourly analysis/forecast cycles & \texttt{sosstsst}, \texttt{sosaline}, \texttt{sossheig} \\
NEMO three-dimensional (validation only; not released) & $40\times311\times225$ per record & Daily & \texttt{votemper}, \texttt{vosaline}, \texttt{vozocrtx}, \texttt{vomecrty} \\
SI3 & $24/72\times311\times225$ & Hourly analysis/forecast cycles & \texttt{siconc}, \texttt{sithic}, \texttt{sivelu}, \texttt{sivelv} \\
WW3 monthly fields & $720$--$744\times225\times311$ & Hourly; monthly files & \texttt{hs}, \texttt{fp}, \texttt{uwnd}, \texttt{vwnd} \\
\bottomrule
\end{tabular}%
}
\end{table}

\subsection{Hindcast validation}
\label{sec:hindcast_results}

All scores in this section were recalculated from the rebuilt matchup arrays using the metric specifications described above. Table~\ref{tab:val_summary} summarizes the observation and observation-based comparisons; model-product intercomparisons follow separately.

\begin{table}[t]
\centering
\caption{Aggregate Borey hindcast validation. Bias is model minus reference; $r$ is Pearson correlation.}
\label{tab:val_summary}
\setlength{\tabcolsep}{4pt}
\resizebox{\textwidth}{!}{%
\begin{tabular}{lllrrrr}
\toprule
Component & Variable & Reference & Period & Bias & RMSE & $r$ \\
\midrule
WRF & 10\,m wind speed & ASCAT & 2019-01 to 2023-08 & $-0.56$\,m\,s$^{-1}$ & $1.84$\,m\,s$^{-1}$ & 0.88 \\
WRF & 10\,m wind speed & ERA5 & 2015-08 to 2022-10 & $+0.00$\,m\,s$^{-1}$ & $1.87$\,m\,s$^{-1}$ & 0.86 \\
WRF & 2\,m air temperature & ERA5 & 2015-08 to 2022-10 & $-0.57$\,K & $1.55$\,K & 0.97 \\
NEMO & Sea surface temperature & OSTIA & 2015-08 to 2023-08 & $-0.63$\,K & $1.60$\,K & 0.93 \\
NEMO & Conservative Temperature profiles & Argo/cruises & 2020-03 to 2023-07 & $-0.50$\,$^{\circ}$C & $0.97$\,$^{\circ}$C & 0.94 \\
NEMO & Absolute Salinity profiles & Argo/cruises & 2020-03 to 2023-07 & $-0.21$\,g\,kg$^{-1}$ & $0.27$\,g\,kg$^{-1}$ & 0.95 \\
SI3 & Sea ice concentration & AMSR2 JAXA & 2015-10 to 2023-08 & $+0.294$ & $0.475$ & 0.57 \\
WW3 & Significant wave height & ESA CCI Sea State L3 & 2015-08 to 2023-08 & $-0.50$\,m & $0.92$\,m & 0.64 \\
WW3 & Significant wave height & CMEMS ALLSAT L4 & 2020-01 to 2023-08 & $-0.61$\,m & $0.99$\,m & 0.62 \\
\bottomrule
\end{tabular}%
}
\end{table}

\subsubsection{Atmosphere}

\begin{figure}[t]
  \centering
  \includegraphics[width=0.95\linewidth]{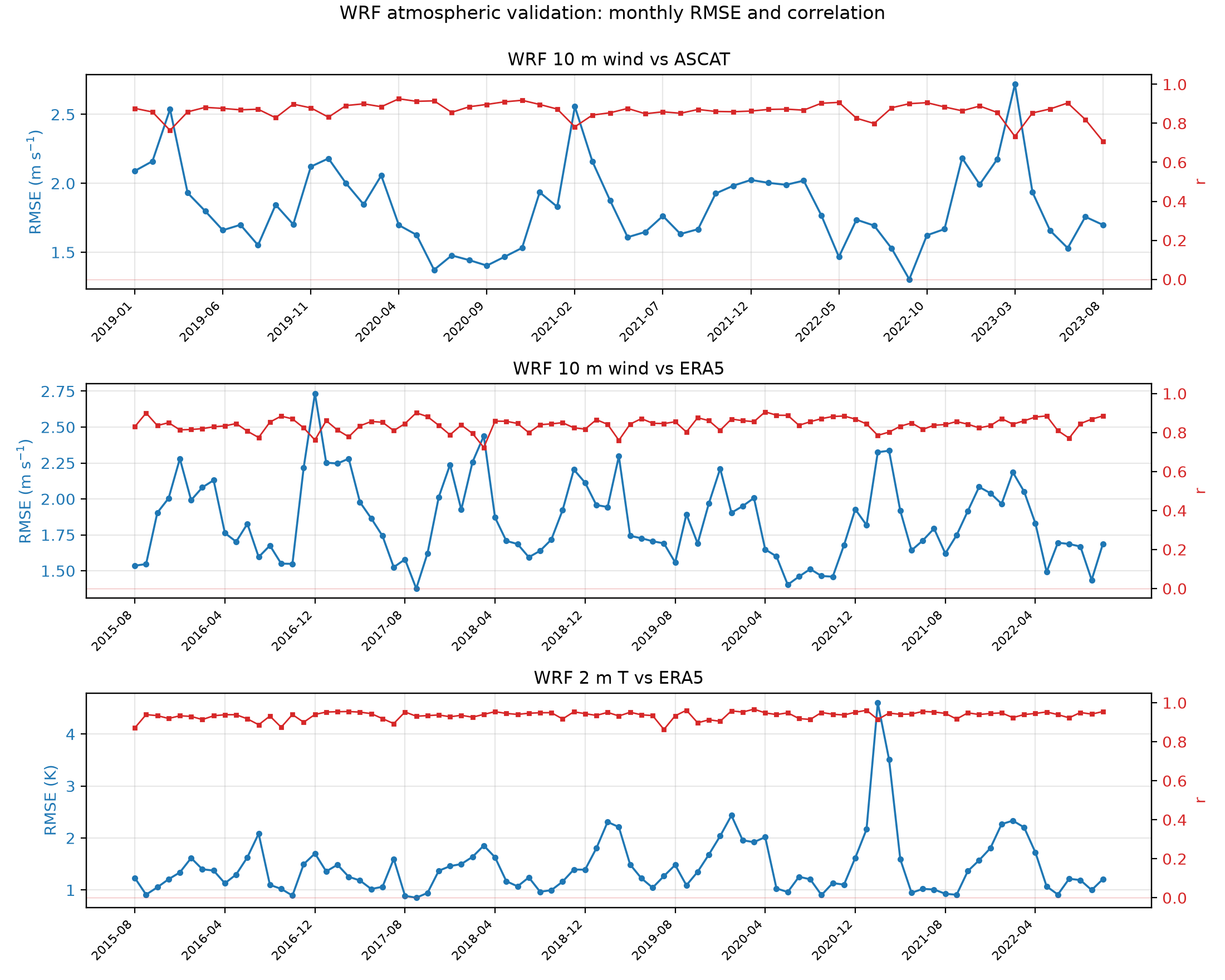}
  \caption{Monthly RMSE and Pearson correlation for the three WRF validation comparisons. RMSE is shown in m\,s$^{-1}$ for wind and K for air temperature.}
  \label{fig:monthly_atmos}
\end{figure}

WRF reproduces most of the temporal variability in marine wind and near-surface air temperature. Wind-speed RMSE is 1.84\,m\,s$^{-1}$ against ASCAT and 1.87\,m\,s$^{-1}$ against ERA5, but the corresponding biases differ: $-0.56$\,m\,s$^{-1}$ against ASCAT and approximately zero against ERA5. This contrast combines regional-model error with differences in reference type and sampling. Air temperature has a $-0.57$\,K bias and $r=0.97$ against ERA5. Monthly errors vary seasonally but show no progressive deterioration over the record (Figure~\ref{fig:monthly_atmos}).

\subsubsection{Ocean}

\begin{figure}[t]
  \centering
  \includegraphics[width=0.95\linewidth]{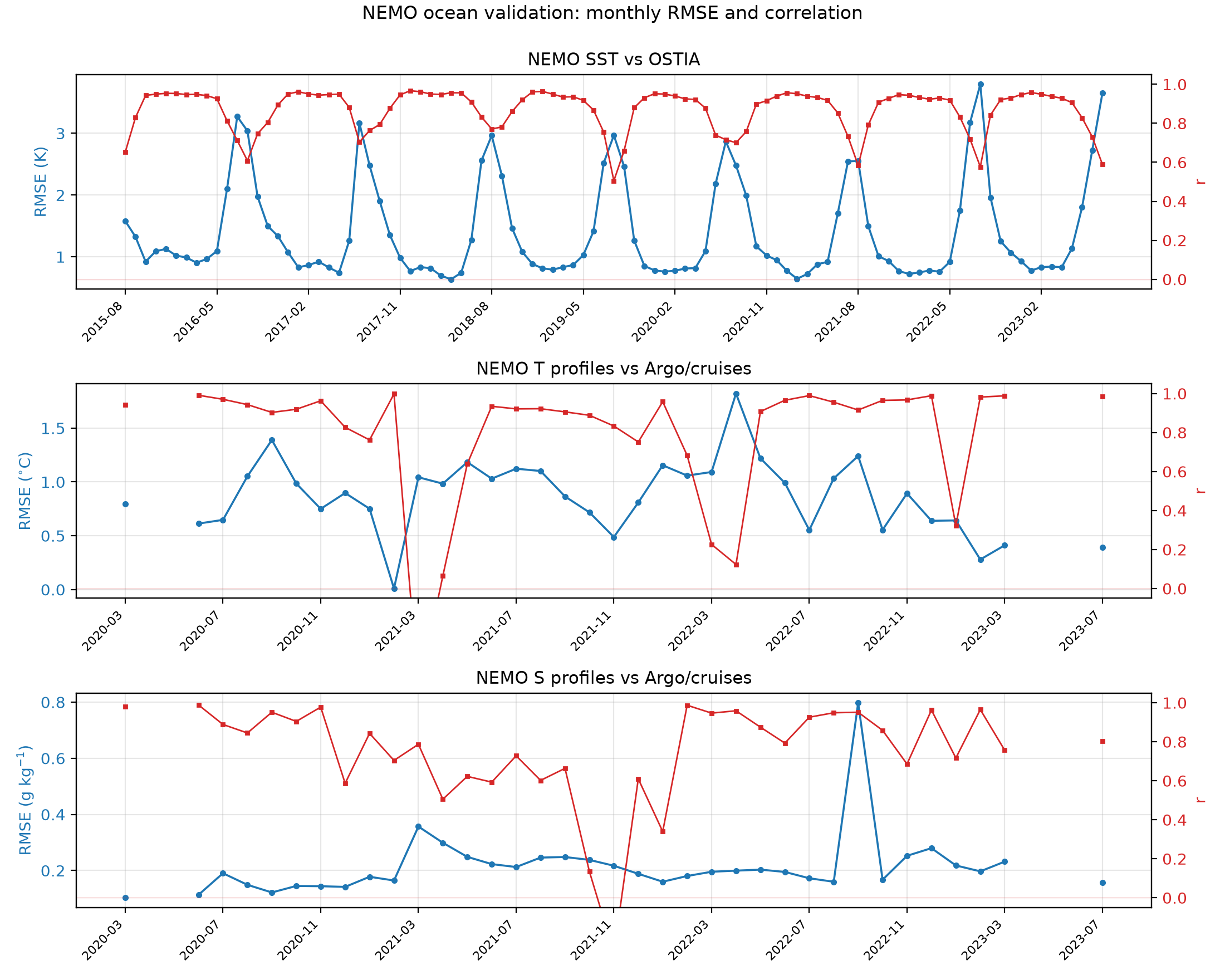}
  \caption{Monthly RMSE and Pearson correlation for NEMO sea surface temperature, Conservative Temperature profiles, and Absolute Salinity profiles. Profile statistics are omitted in months without sufficient observations; Absolute Salinity RMSE is in g\,kg$^{-1}$.}
  \label{fig:monthly_ocean}
\end{figure}

Sea surface temperature has a persistent cold bias and higher RMSE during the melt season (Figure~\ref{fig:monthly_ocean}). The profile comparison also shows a cold Conservative Temperature bias of $-0.50$\,$^{\circ}$C and a fresh Absolute Salinity bias of $-0.21$\,g\,kg$^{-1}$, with correlations of 0.94 and 0.95, respectively. Because profile coverage is sparse and seasonally uneven, the aggregate scores are more representative than individual monthly values. The coincident seasonal SST and sea ice errors are an association; this validation does not establish a causal mechanism.

\subsubsection{Sea ice}
\label{sec:val_ice}

\begin{figure}[t]
  \centering
  \includegraphics[width=0.95\linewidth]{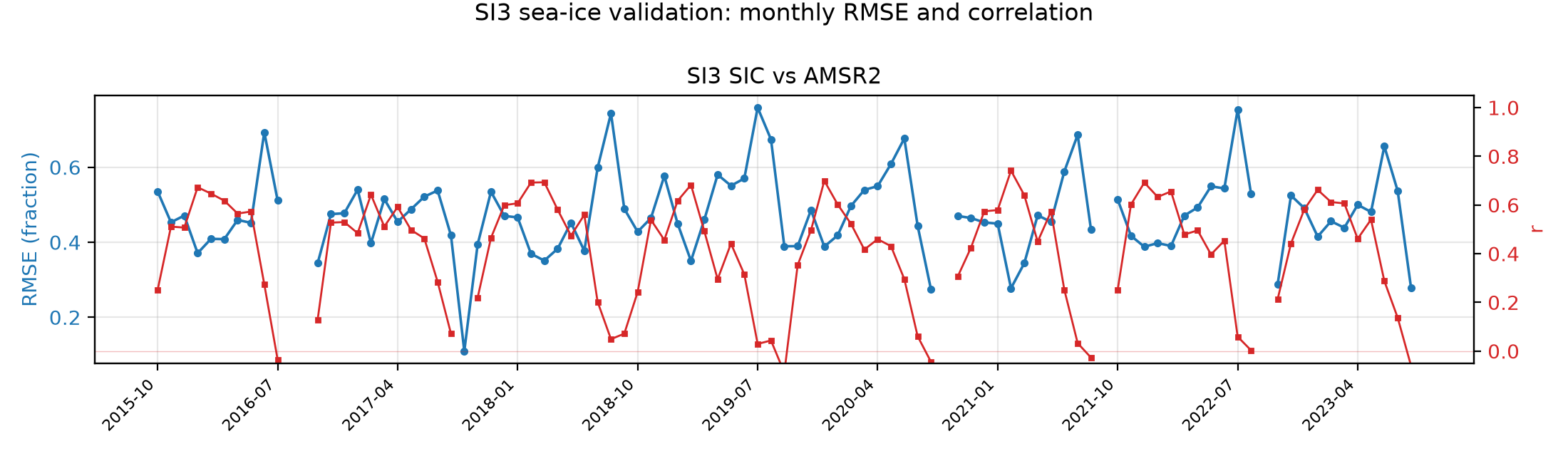}
  \caption{Monthly RMSE and Pearson correlation for SI3 sea ice concentration against AMSR2 JAXA.}
  \label{fig:monthly_ice}
\end{figure}

Sea ice concentration has a positive mean bias of $+0.294$ and RMSE of 0.475 against AMSR2. At the 0.15 concentration threshold, the accumulated ice-state disagreement is $2.04\times10^{7}$ cell-time cases; 96\,\% are model-only ice and 4\,\% are reference-only ice. This diagnostic is a count rather than an area-weighted integrated extent error. RMSE increases and correlation decreases during summer retreat (Figure~\ref{fig:monthly_ice}), consistent with the strongest disagreement near the seasonal ice edge. Thickness and drift are assessed only against contextual products in Section~\ref{sec:context_intercomparisons}.

\subsubsection{Waves}

\begin{figure}[t]
  \centering
  \includegraphics[width=0.95\linewidth]{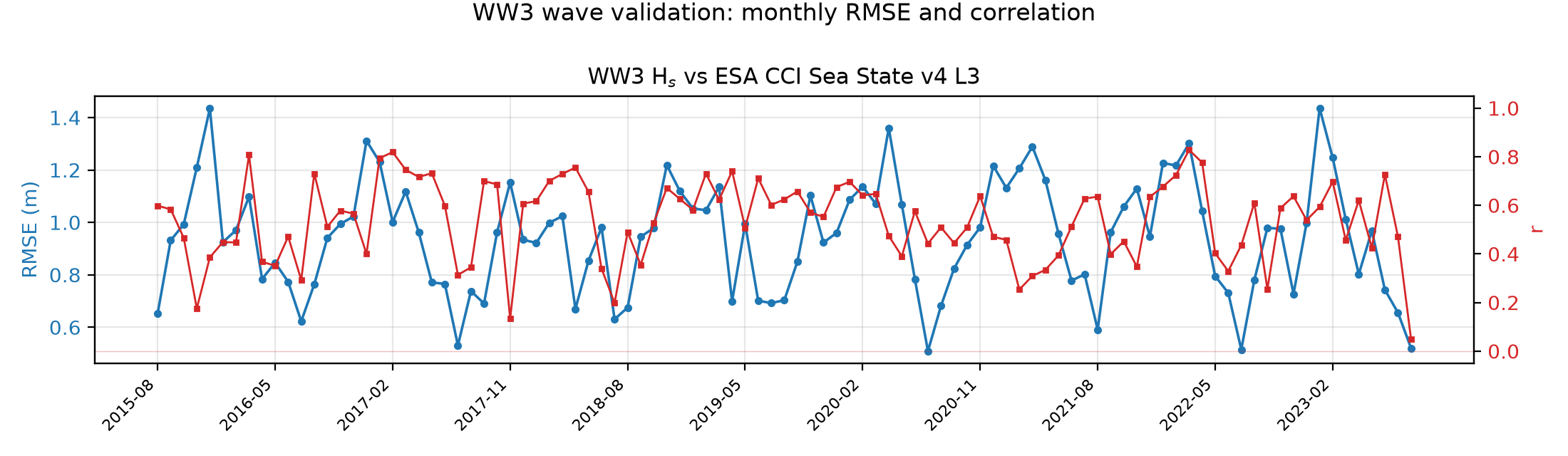}
  \caption{Monthly RMSE and Pearson correlation for WW3 significant wave height against the ESA CCI Sea State version 4 Level-3 hourly 0.25$^{\circ}$ derivative.}
  \label{fig:monthly_wave}
\end{figure}

Against the local hourly 0.25$^{\circ}$ CCI Sea State version 4 Level-3 derivative, WW3 significant wave height has a mean bias of $-0.50$\,m, MAE of 0.65\,m, RMSE of 0.92\,m, and $r=0.64$. The bias is negative in every evaluated month, while RMSE is generally higher in winter and early spring and lower in summer (Figure~\ref{fig:monthly_wave}). These scores represent sparse, observation-occupied CCI cells after quality control and ice screening, not a spatially complete domain mean.

Against the CMEMS ALLSAT Level-4 representation, 73,770 matched pairs yield a mean bias of $-0.61$\,m, MAE of $0.73$\,m, RMSE of $0.99$\,m, and $r=0.62$. Relative to the CCI comparison, the ALLSAT result has a slightly more negative bias, nearly identical RMSE, and slightly lower correlation. Because this comparison samples a coarser gridded representation of overlapping multi-mission altimeter observations, the differences are interpreted as sensitivity to reference processing and sampling rather than as an independent replication of the CCI validation.

\subsection{Contextual product intercomparisons}
\label{sec:context_intercomparisons}

TOPAZ5 and AVISO provide context for variables without a fully independent, domain-wide observational reference (Table~\ref{tab:baseline_summary}). TOPAZ5 is an assimilative analysis and forecast product, while AVISO resolves the geostrophic component of surface flow. These comparisons are therefore not treated as equivalent to the observation-based rows in Table~\ref{tab:val_summary}.

\begin{table}[t]
\centering
\caption{Borey intercomparisons with TOPAZ5 and AVISO. Bias, RMSE, and $r$ use matched speed magnitudes for drift and currents.}
\label{tab:baseline_summary}
\setlength{\tabcolsep}{4pt}
\resizebox{\textwidth}{!}{%
\begin{tabular}{lllrrrrr}
\toprule
Component & Variable & Reference & Period & Matchups & Bias & RMSE & $r$ \\
\midrule
SI3 & Sea ice concentration & TOPAZ5 analysis+forecast & 2021-07 to 2023-08 & 10\,573\,466 & $+0.220$ & $0.382$ & 0.69 \\
SI3 & Sea ice thickness & TOPAZ5 analysis+forecast & 2021-07 to 2023-08 & 9\,246\,232 & $+0.70$\,m & $0.89$\,m & 0.73 \\
SI3 & Daily drift speed & TOPAZ5 analysis+forecast & 2021-07 to 2023-08 & 9\,177\,554 & $+0.019$\,m\,s$^{-1}$ & $0.099$\,m\,s$^{-1}$ & 0.47 \\
NEMO & Surface current speed & TOPAZ5 analysis+forecast & 2021-07 to 2023-08 & 24\,021\,943 & $-0.024$\,m\,s$^{-1}$ & $0.073$\,m\,s$^{-1}$ & 0.44 \\
NEMO & Surface current speed (total) & AVISO L4 geostrophic & 2015-08 to 2023-08 & 96\,381\,992 & $+0.014$\,m\,s$^{-1}$ & $0.063$\,m\,s$^{-1}$ & 0.43 \\
NEMO & Surface current speed (geostrophic) & AVISO L4 geostrophic & 2015-08 to 2023-08 & 28\,755\,197 & $+0.009$\,m\,s$^{-1}$ & $0.058$\,m\,s$^{-1}$ & 0.49 \\
\bottomrule
\end{tabular}%
}
\end{table}

The sea ice concentration bias has the same sign against TOPAZ5 and AMSR2. SI3 thickness exceeds TOPAZ5 by 0.70\,m on average, but this is not an independent thickness validation because TOPAZ5 assimilates satellite thickness information. The NEMO geostrophic-current comparison has a slightly lower RMSE and higher correlation than the total-current comparison with AVISO. Different matchup subsets and the absence of ageostrophic flow in AVISO prevent attributing that difference to a single cause.

\begin{table}[t]
\centering
\caption{Direction-aware statistics for surface currents and sea ice drift. RMSVE is root-mean-square vector error; $|\rho_c|$ and $\arg(\rho_c)$ are the magnitude and phase of complex vector correlation.}
\label{tab:vector_summary}
\setlength{\tabcolsep}{4pt}
\resizebox{\textwidth}{!}{%
\begin{tabular}{lllrrrrrr}
\toprule
Component & Variable & Reference & Matchups & $|\rho_c|$ & $\arg(\rho_c)$ & RMSVE & Mean absolute angle error & Signed angle bias \\
\midrule
NEMO & Surface currents (total) & AVISO L4 geostrophic & 96\,381\,992 & 0.44 & $-0.8^{\circ}$ & $0.090$\,m\,s$^{-1}$ & $65.8^{\circ}$ & $-0.6^{\circ}$ \\
NEMO & Surface currents (geostrophic) & AVISO L4 geostrophic & 28\,755\,197 & 0.49 & $-0.2^{\circ}$ & $0.082$\,m\,s$^{-1}$ & $61.4^{\circ}$ & $-1.2^{\circ}$ \\
NEMO & Surface currents (total) & TOPAZ5 analysis+forecast & 24\,021\,943 & 0.55 & $-0.9^{\circ}$ & $0.102$\,m\,s$^{-1}$ & $54.0^{\circ}$ & $-0.3^{\circ}$ \\
SI3 & Daily drift & TOPAZ5 analysis+forecast & 9\,177\,554 & 0.64 & $+8.0^{\circ}$ & $0.113$\,m\,s$^{-1}$ & $39.0^{\circ}$ & $+7.4^{\circ}$ \\
\bottomrule
\end{tabular}%
}
\end{table}

\begin{figure}[t]
  \centering
  \includegraphics[width=0.95\linewidth]{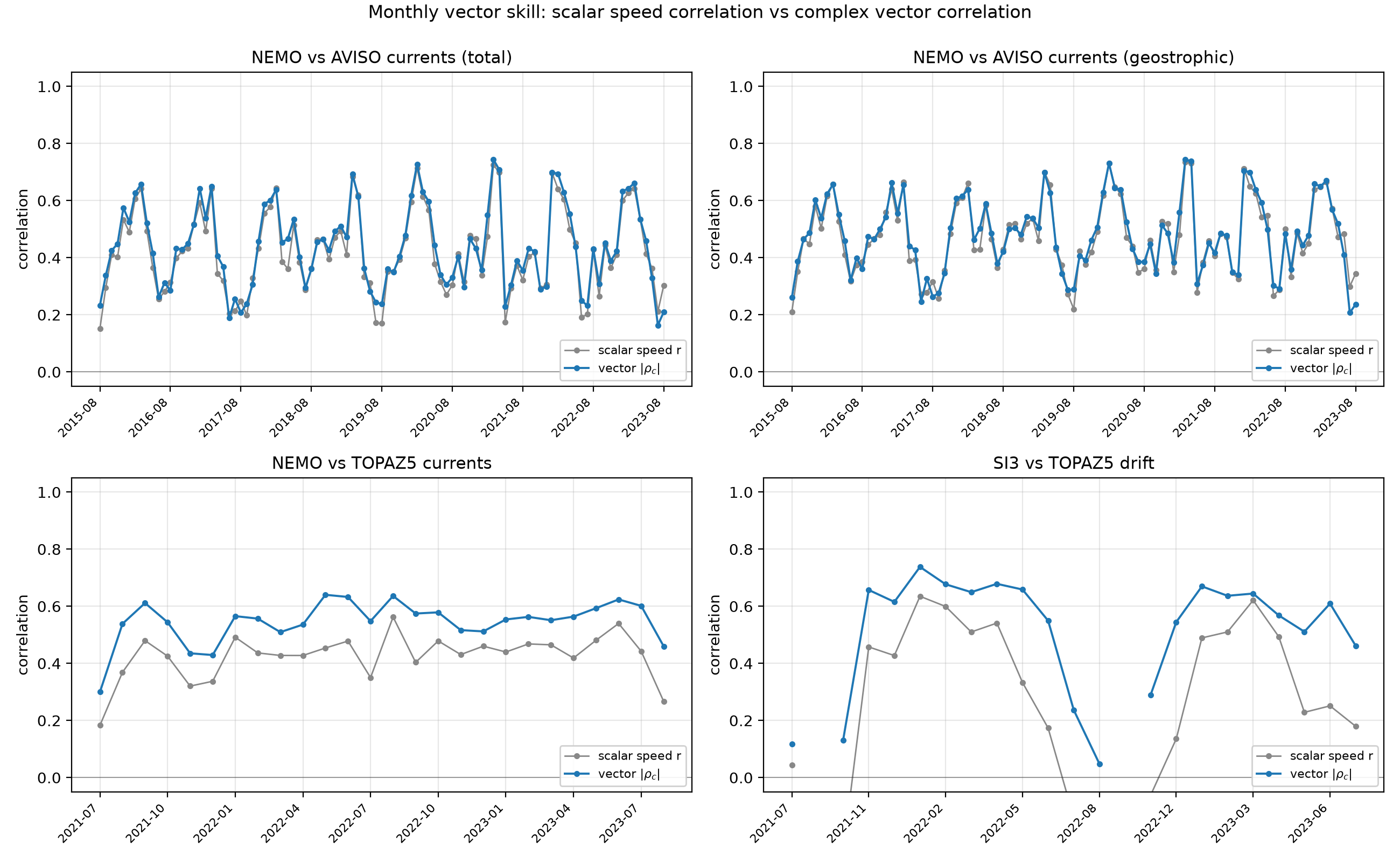}
  \caption{Monthly vector skill for surface currents and sea ice drift. Each panel compares the scalar speed correlation (grey) with the magnitude of the complex vector correlation $|\rho_c|$ (blue) for NEMO currents against AVISO (total and geostrophic), NEMO currents against TOPAZ5, and SI3 drift against TOPAZ5.}
  \label{fig:vector_metrics}
\end{figure}

\begin{figure}[t]
  \centering
  \includegraphics[width=0.95\linewidth]{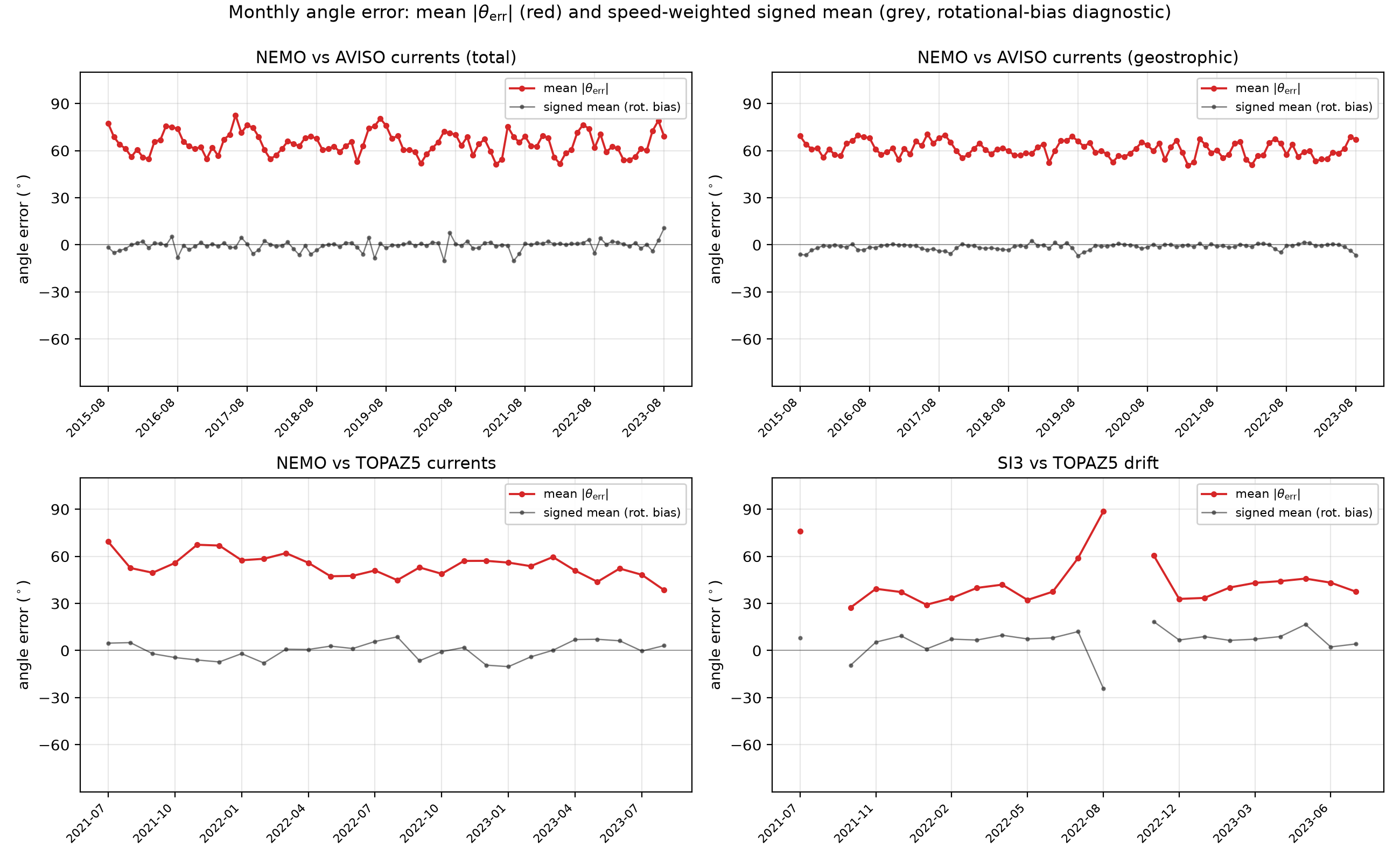}
  \caption{Monthly direction error for the comparisons in Figure~\ref{fig:vector_metrics}: the mean absolute angle error $\langle|\theta_{\mathrm{err}}|\rangle$ (red) and the speed-weighted signed mean angle error (grey), the latter a rotational-bias diagnostic. Positive signed values indicate a mean counter-clockwise rotation of the modeled vectors relative to the reference.}
  \label{fig:vector_angles}
\end{figure}

In the TOPAZ5 comparison, SI3 represents drift direction more consistently than drift-speed variability: complex vector correlation exceeds scalar speed correlation in most months (Figure~\ref{fig:vector_metrics}). Direction errors are larger for surface currents, especially when total NEMO current is compared with geostrophic AVISO current. The large mean absolute angle error contrasts with a small speed-weighted signed rotational bias (Figure~\ref{fig:vector_angles}), indicating broad directional scatter without a comparably large preferred rotation. These benchmarks constrain the scales and current components that can be interpreted; they do not diagnose the full regional circulation.

\subsection{Forecast diagnostics}
\label{sec:forecast_results}

\begin{figure}[t]
  \centering
  \includegraphics[width=0.95\linewidth]{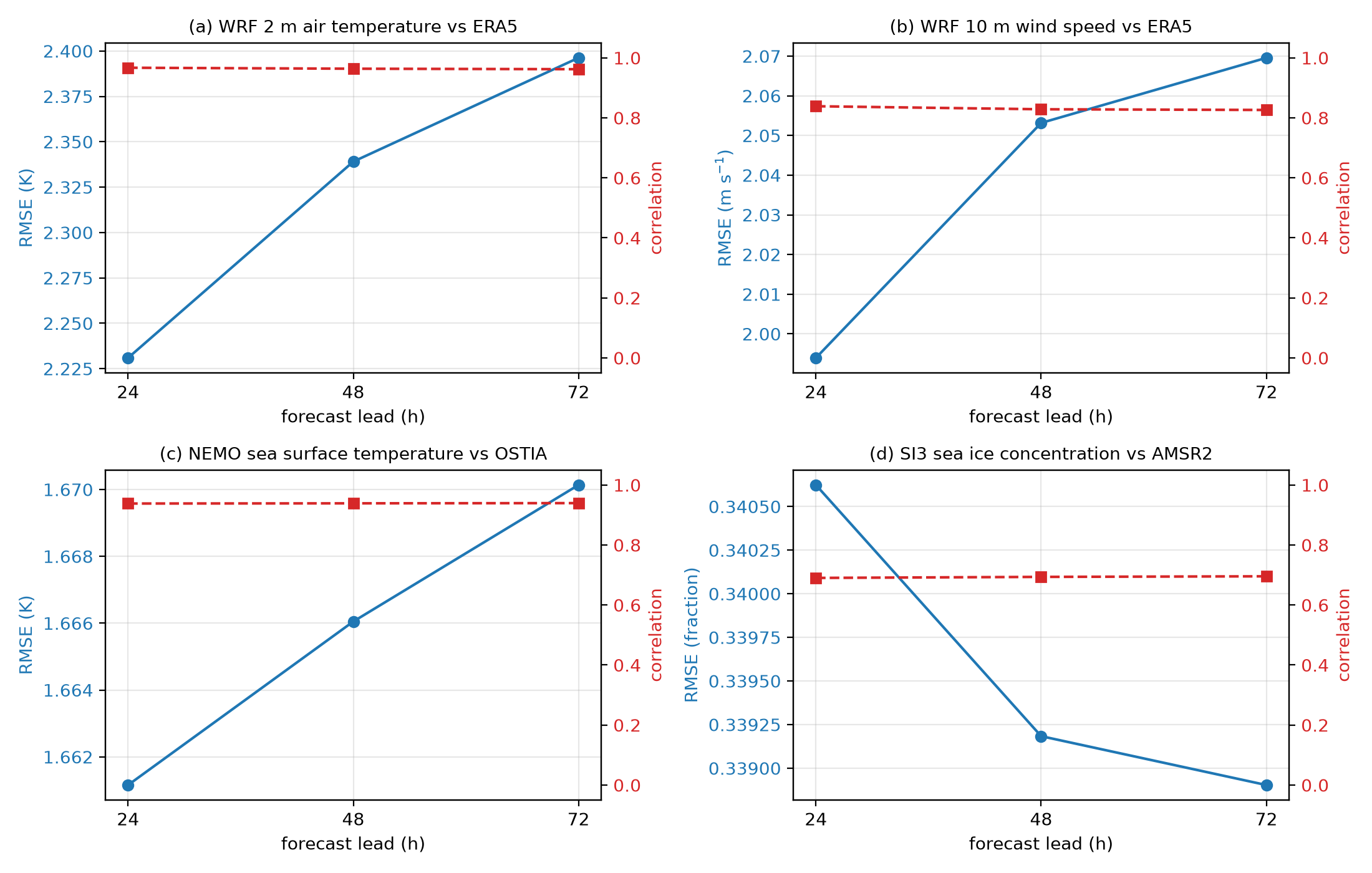}
  \caption{Borey operational forecast skill versus lead time (24, 48, 72\,h) for (a) 2\,m air temperature and (b) 10\,m wind speed against ERA5, (c) sea surface temperature against OSTIA, and (d) sea ice concentration against AMSR2 JAXA. Each panel shows RMSE (blue, left axis) and Pearson correlation (red, right axis) on common-cycle cohorts of complete 00 UTC cycles from July 2025 through February 2026.}
  \label{fig:forecast_leads}
\end{figure}

The operational chain was evaluated using complete 00 UTC cycles from July 2025 through February 2026 and the common-cycle approach defined in Section~\ref{sec:forecast_validation}. Table~\ref{tab:forecast_summary} and Figure~\ref{fig:forecast_leads} summarize the four completed WRF, NEMO, and SI3 comparisons.

Forecast skill changes only modestly across the 72-hour window. Two-metre air-temperature RMSE against ERA5 increases from 2.23\,K at 24\,h to 2.40\,K at 72\,h, and 10\,m wind-speed RMSE from 1.99 to 2.07\,m\,s$^{-1}$; correlation remains near 0.97 and 0.83, respectively. Sea surface temperature against OSTIA is nearly flat across leads, as is sea ice concentration against AMSR2, with ice-state disagreement dominated by model-only ice. Bias signs are consistent with the corresponding hindcast comparisons: cold air and sea surface temperatures, near-zero wind bias, and excess ice concentration. This weak lead dependence indicates limited error growth within the evaluated window, but it does not identify whether persistent errors arise from forcing, initialization, or model configuration. Because the recent forecast cohort differs from the multi-year hindcast period, their absolute scores are not directly comparable. Wave and surface-current forecast results are omitted until their validation is complete.

\begin{table}[t]
\centering
\caption{Borey operational forecast skill by lead time, evaluated on common-cycle cohorts of complete 00 UTC cycles from July 2025 through February 2026. Bias is model minus reference; $r$ is Pearson correlation. Cohort sizes: 2\,m temperature 137 cycles, 10\,m wind 138, sea surface temperature 208, sea ice concentration 107.}
\label{tab:forecast_summary}
\setlength{\tabcolsep}{4pt}
\resizebox{\textwidth}{!}{%
\begin{tabular}{lllrrrr}
\toprule
Component & Variable & Reference & Lead (h) & Bias & RMSE & $r$ \\
\midrule
WRF & 2\,m air temperature & ERA5 & 24 & $-0.72$\,K & $2.23$\,K & 0.97 \\
     &                      &      & 48 & $-0.70$\,K & $2.34$\,K & 0.96 \\
     &                      &      & 72 & $-0.68$\,K & $2.40$\,K & 0.96 \\
\addlinespace
WRF & 10\,m wind speed & ERA5 & 24 & $+0.09$\,m\,s$^{-1}$ & $1.99$\,m\,s$^{-1}$ & 0.84 \\
     &                  &      & 48 & $+0.10$\,m\,s$^{-1}$ & $2.05$\,m\,s$^{-1}$ & 0.83 \\
     &                  &      & 72 & $+0.12$\,m\,s$^{-1}$ & $2.07$\,m\,s$^{-1}$ & 0.83 \\
\addlinespace
NEMO & Sea surface temperature & OSTIA & 24 & $-0.90$\,K & $1.66$\,K & 0.94 \\
      &                         &       & 48 & $-0.91$\,K & $1.67$\,K & 0.94 \\
      &                         &       & 72 & $-0.92$\,K & $1.67$\,K & 0.94 \\
\addlinespace
SI3 & Sea ice concentration & AMSR2 JAXA & 24 & $+0.103$ & $0.341$ & 0.69 \\
     &                       &            & 48 & $+0.103$ & $0.339$ & 0.69 \\
     &                       &            & 72 & $+0.104$ & $0.339$ & 0.70 \\
\bottomrule
\end{tabular}%
}
\end{table}

\section{Discussion}

Borey links an operational regional forecast system to a multicomponent hindcast with a continuous ocean--sea ice trajectory and documented atmospheric and wave forcing. The absence of assimilation increments supports process analysis, model development, and downstream experiments, while also allowing systematic regional errors to persist. Borey should therefore complement rather than replace observation-constrained products.

The clearest coupled seasonal pattern is a cold sea surface temperature bias alongside excessive sea ice concentration and occurrence during retreat. Thermodynamics, atmospheric forcing, and initial or boundary conditions could all contribute, but the present comparisons cannot separate them. The positive thickness difference relative to TOPAZ5 provides context only; it is not independent evidence that excessive thickness causes the concentration error.

Atmospheric variability is represented with high correlation against the retained references, while the different ASCAT and ERA5 wind biases demonstrate the importance of reference type and sampling. Current and drift intercomparisons are similarly constrained by reference physics and effective resolution. The CCI assessment identifies a persistent negative WW3 significant-wave-height bias, with larger errors in winter and early spring. Its sparse, quality-controlled, open-water sampling means the aggregate score is not a domain-wide wave climatology.

Across the evaluated operational cycles, air-temperature and wind errors grow only slightly with lead, while sea surface temperature and sea ice concentration scores are nearly unchanged. The consistency of bias signs with the hindcast suggests a persistent component to short-range error, making the hindcast useful for bias characterization. The experiment does not, however, isolate model formulation from forcing or initialization, and its recent common-cycle cohorts are too limited to establish seasonal forecast performance.

Priorities for future work are diagnosis and reduction of the persistent ocean, sea ice, and wave biases; evaluation over longer operational forecast cohorts; and completion of wave and surface-current forecast validation. An assimilative Borey variant and two-way atmosphere--ocean--sea ice--wave coupling remain separate development directions.

\section{Conclusions}
Borey combines WRF, NEMO-SI3, and WW3 in an operational atmosphere--ocean--sea ice--wave system for the Barents and Kara Seas. It produces daily forecasts to 72 hours and an aligned hourly surface record on approximately 3--6\,km grids. The accompanying hindcast extends from August 2015 through August 2023 and provides the long baseline needed to interpret current operational behavior.

The validation shows that Borey captures much of the observed variability in near-surface atmospheric conditions and ocean temperature. Its principal limitations are persistent cold sea surface temperature, excessive sea ice during seasonal retreat, and underestimated significant wave height. In the recent operational cohorts, WRF, NEMO, and SI3 skill changes only modestly from the first to the third forecast day. This stability makes bias reduction a clear priority, while the available diagnostics do not yet distinguish among errors introduced by model formulation, forcing, and initialization.

The public dataset will intentionally contain surface fields only. Three-dimensional NEMO fields were used to validate subsurface temperature and salinity but are not required for the intended archive and will not be released. Together with native grids and configuration and provenance documentation, the complete surface dataset and operational forecasts provide a basis for regional marine and sea ice analysis, model development, and carefully evaluated data-driven forecasting and data-assimilation research. Borey should be used alongside observation-constrained products, with the documented biases considered in any application.

\section*{Author Contributions}

Vasily Ivanov designed and developed the original pipeline architecture, led the scientific validation, performed ocean and sea ice model tuning, and wrote most of the manuscript. Polina Verezemskaya developed the NEMO configuration, implemented its code and initial parameterization, and contributed specialist expertise in ocean and sea ice modeling. Alexander Gavrikov developed and extensively tuned the WRF configuration, conducted configuration experiments, and contributed specialist expertise in atmospheric modeling with WRF. Vitaliy Sharmar developed the WW3 configuration and contributed specialist expertise in wave modeling with WW3. Mikhail Krinitskiy contributed to scientific validation, critical evaluation of the results, manuscript review, and model-development tooling. Timofey Grigoryev served as a co-lead developer with Vasily Ivanov and implemented substantial parts of the system code. Vladimir Vanovskiy conceived the combined multicomponent configuration, led the research, and organized the overall project. Evgeny Burnaev provided strategic scientific guidance and critical review and helped establish the project's research priorities.

\section*{Conflict of Interest Statement}

The authors declare that the research was conducted in the absence of any commercial or financial relationships that could be construed as a potential conflict of interest.

\section*{Data Availability Statement}

The Borey hindcast dataset is being prepared for public release. It will contain the complete hourly surface dataset, native grids and masks, and documentation of the model configurations and production provenance. Three-dimensional NEMO fields used for profile validation are outside the release scope and will not be distributed. Repository information, a persistent identifier, format and metadata documentation, and licensing terms will accompany the deposited version. External forcing and validation products are not redistributed and must be obtained from their original providers.

The planned public release does not include the model, workflow, or validation source code.

\bibliographystyle{Frontiers-Harvard}
\bibliography{references}

\end{document}